\documentclass[aps,epsf,twocolumn,showpacs]{revtex4}
\usepackage{amsmath}
\usepackage{epsfig}

\begin{document}

\title{Universality of the Ising and the $S=1$ model on Archimedean
lattices: A Monte Carlo determination}

\author{A. Malakis$^1$}

\author{G. Gulpinar$^2$}

\author{Y. Karaaslan$^2$}

\author{T. Papakonstantinou$^1$}

\author{G. Aslan$^2$}

\affiliation{$^1$Department of Physics, Section of Solid State
Physics, University of Athens, Panepistimiopolis, GR 15784
Zografos, Athens, Greece}

\affiliation{$^2$Department of Physics, Dokuz Eylul University,
Buca 35160, Izmir, Turkey}

\date{\today}

\begin{abstract}
The Ising model $S=1/2$ and the $S=1$ model are studied by
efficient Monte Carlo schemes on the (3,4,6,4) and the (3,3,3,3,6)
Archimedean lattices. The algorithms used, a hybrid
Metropolis-Wolff algorithm and a parallel tempering protocol, are
briefly described and compared with the simple Metropolis
algorithm. Accurate Monte Carlo data are produced at the exact
critical temperatures of the Ising model for these lattices. Their
finite-size analysis provide, with high accuracy, all critical
exponents which, as expected, are the same with the well known 2d
Ising model exact values. A detailed finite-size scaling analysis
of our Monte Carlo data for the $S=1$ model on the same lattices
provides very clear evidence that this model obeys, also very
well, the 2d Ising model critical exponents. As a result, we find
that recent Monte Carlo simulations and attempts to define
effective dimensionality for the $S=1$ model on these lattices are
misleading. Accurate estimates are obtained for the critical
amplitudes of the logarithmic expansions of the specific heat for
both models on the two Archimedean lattices.

\end{abstract}

 \maketitle

\section{Introduction}
\label{sec:1}

The Ising model, and several of its generalizations, have been of
central importance in the development of the theory of phase
transitions and the formulation of the universality
hypothesis~\cite{F66,Griff70}. According to this hypothesis all
critical systems with the same dimensionality, the same symmetry
of the ordered phase, and the same number of order parameters are
expected to share the same set of critical exponents. For the 2d
Ising model (square and some other lattices) all critical
exponents are known exactly~\cite{O44,K49,Y52,B82}. These
exponents are expected to be obeyed by the Ising model on all
two-dimensional (2d) lattices and also by all other models, which
according to the above hypothesis are expected to belong to the
same universality class.

In several cases, this expectation has been verified either by
exact analytic solutions~\cite{B82}, or, with impressive
accuracy~\cite{LG80,ferrenberg91,BLH95,MHAS10}, using Monte Carlo
(MC) simulations and the theory of finite-size scaling
(FSS)~\cite{F71,VPrvm90,KB92}. Furthermore, the issue of explicit
finite-size expansion of the main thermodynamic functions, or
their accurate numerical estimation, has been considered in
several cases and is of substantial significance. Often this
provides a tool to improve accuracy of the MC estimation of
critical exponents, especially in cases complicated by the
presence of logarithmic corrections~\cite{kenna}. In particular,
the studies of critical amplitudes of the specific heat
expansions~\cite{Ferdin69,malakis04,Izmail02,Izmail03,salas01,SchJanke10}
and the studies of universal and non-universal features of certain
combinations of critical amplitudes of the order
parameter~\cite{SchJanke10,Blote93,Selke05,SalasSokal98} are very
interesting topics. One of the well-known generalizations of the
Ising model is the $S=1$ model studied in this paper. Here, we
will present a careful MC study verifying, with high accuracy, the
universality hypothesis of the two models, and we will also report
results related to the logarithmic expansions of the specific
heat, for two different 2d lattices known as Archimedean lattices.
Our motivation is also to test a recent MC study~\cite{Lima10},
that erroneously resulted in an attempt to define and estimate
effective dimensionality for the $S=1$ model on these lattices.

\begin{figure}[htbp]
\includegraphics*[width=9 cm]{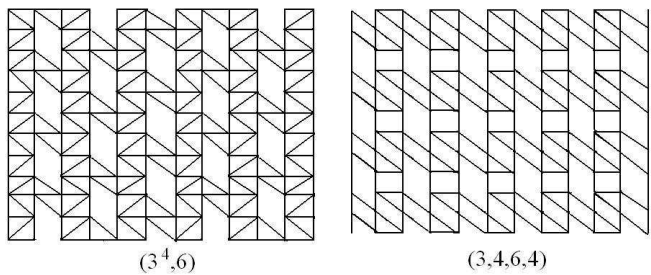}
\caption{\label{Fig:1} Structure of the ($3^{4}, 6$) and (3,4,6,4)
Archimedean lattices (ALs). For the Monte Carlo simulations we
used periodic boundary conditions.}
\end{figure}

An Archimedean lattice is a graph of a regular tiling of the plane
whose all corners are equivalent and are shared by the same set of
polygons. Thus, we may denote each Archimedean lattice by a set of
integers ($p_1,p_2,...$) indicating, in cyclic order, the polygons
meeting at a given vertex. As an example, the square lattice is
the Archimedean lattice denoted by ($4^4$). There exists eleven 2d
Archimedean lattices. In addition, they have dual lattices, three
of which are Archimedean, the other eight are entitled Laves
lattices~\cite{Uppsala}. Suding and Ziff presented precise
thresholds for site percolation on eight Archimedean lattices
determined by the hull-walk gradient-percolation simulation
method~\cite{Ziff99}. Rigorous bounds for the bond percolation
critical probability are determined for three Archimedean lattices
by Wierman~\cite{Wierman02}. In addition, Scullard and  Ziff
showed that the exact determination of the bond percolation
threshold for the Martini lattice can be used to provide
approximations to the Kagome and ($3,12^{2}$)
lattices~\cite{Scullard06}. As illustrated by Suding and Ziff, the
Archimedean lattices can be transformed to a square array of
$N=L^{2}$ vertices and then apply periodic boundary conditions.
For the computer simulation of the two models studied in this
paper, we also apply periodic boundary conditions (PBC) and use
their representation as illustrated in Fig. 3 of their
paper~\cite{Ziff99}. The two Archimedean lattices (ALs), used in
our study, are the ($3^{4}, 6$) and (3,4,6,4) lattices illustrated
in Fig.~\ref{Fig:1}.

In 2005, Malarz \emph{et al.} evaluated critical temperatures for
the ferro-paramagnetic transition in the Ising model for five
Archimedean lattices, utilizing  Monte Carlo (MC) simulations
\cite{Malarz05}. Then ($3, 4, 6, 4$) and ($3^{4}, 6$) Archimedean
lattices of the majority-vote model with noise are considered and
studied through MC simulations and the order/disorder phase
transition is observed in this study~\cite{Lima06}. Later,
Krawczyk \emph{et al.} investigated the magnetic properties of a
2d periodic structure which is topologically equivalent to the
Archimedean ($3, 12^{2}$) lattice with Ising spins and
antiferromagnetic nearest-neighbor interaction~\cite{Krawcyzk04}.
In 2010 Codello determined the exact Curie temperature for all 2d
Archimedean lattices by making use of the Feynman-Vdovichenko
combinatorial approach to the two dimensional Ising
model~\cite{Codelo10}. Recently, Lima \emph{et al.} studied the
critical properties of the Ising $S = 1/2$ and $S = 1$ model on
($3, 4, 6, 4$) and ($3^{4}, 6$) Archimedean lattices with MC
simulations~\cite{Lima10}. In this study they claimed that the
Ising $S = 1$ model on ($3, 4, 6, 4$) and ($3^{4}, 6$) AL exhibits
a second-order phase transition with critical exponents which do
not fall into universality class of the square lattice Ising
$S=1/2$ model.

The outline of the paper is as follows. In Sec.~\ref{sec:2}, after
the definition of the models, we present details of the MC
schemes, used in this paper, and compare their performance with
the simple Metropolis algorithm by illustrating the behavior of
moving averages for the order parameter of the systems. In
Sec.~\ref{sec:3} we present accurate MC data obtained at the exact
critical temperatures~\cite{Codelo10} for the Ising model on the
(3,4,6,4) and ($3^4$,6) ALs. Their FSS behavior yields with high
accuracy all the expected critical properties of the universality
class of the 2d Ising model. The logarithmic expansions of the
corresponding specific heat on the two ALs are also studied. In
Sec.~\ref{sec:4} we estimate the critical temperatures and the
corresponding critical properties of the $S=1$ model on the
(3,4,6,4) and ($3^4$,6) ALs with periodic boundary conditions.
Estimates are given for the corresponding critical amplitudes of
the specific heat expansions. This is carried out by observing the
FSS behavior of several finite size anomalies of the AL systems.
Furthermore, the finite size scaling behavior at the accurately
estimated critical temperatures is discussed. In the conclusions,
summarized in Sec.~\ref{sec:4}, we comment on previous
observations concerning the critical behavior of the $S=1$ model
on these lattices.

\section{Definition of the Models, Monte Carlo schemes and FSS approach}
\label{sec:2}

The (zero-field) Ising model is defined by the Hamiltonian
\begin{equation}
\label{eq:1}
 H=-J\sum_{\langle ij \rangle}s_{i}s_{j},
\end{equation}
with spin variables $s_{i}$ taking on the values $-1$, or $+1$. As
usual $\langle ij \rangle$ indicates summation over all
nearest-neighbor pairs of sites, and $J>0$ for ferromagnetic
exchange interaction. There is a variety of possible
generalizations of the Ising model. Keeping only nearest-neighbor
interactions one can generalize to a $S=1$ model including up to
five interaction constants~\cite{collins88}. This is a rich model
describing several phase transitions, critical and multicritical
phenomena with a wide range of physical applications. Special
cases of this most general model are the well known and
extensively studied Blume-Capel (BC) model~\cite{blume66,capel66}
and also the Blume-Emery-Griffiths (BEG) model~\cite{blume71}. For
our purposes, it suffices to introduce only the above mentioned
generalization known as the BC model~\cite{blume66,capel66}. It is
defined by introducing spin variables $s_{i}$ that take on the
values $-1, 0$, or $+1$, and a crystal field coupling $\Delta$, so
that the Hamiltonian is given by
\begin{equation}
\label{eq:2} H=-J\sum_{\langle ij \rangle
}s_{i}s_{j}+\Delta\sum_{i}s_{i}^{2}.
\end{equation}

This model is of particular importance for the theory of phase
transitions and critical phenomena, since as is well known its
phase diagram consists of a segment of continuous Ising-like
transitions at high temperatures and for low values of the crystal
field which ends at a tricritical point, where it is joined with a
second segment of first-order transitions between
($\Delta_{t},T_{t}$) and ($\Delta_{0},T=0$). The BC model has been
analyzed, besides the original mean-field
theory~\cite{blume66,capel66}, by a variety of approximations and
numerical approaches, in both 2d and 3d. These include the real
space renormalization group~\cite{WortisBerker1976}, MC
simulations~\cite{landau72}, and MC renormalization-group
calculations ~\cite{Swedsenlandau86}, $\epsilon$-expansion
renormalization groups~\cite{stephen73}, high- and low-temperature
series calculations~\cite{fox73}, a phenomenological FSS analysis
using a strip geometry~\cite{nightingale82,beale86}, and MC
simulations~\cite{malakis09,jain80,landau81,care93,deserno97,silva06}.
In particular, the 2d (mainly the square BC model) has been
extensively studied and there is no doubt today that the
continuous Ising-like transitions, along its second-order segment,
obey the same critical properties with the 2d Ising model.
Recently, a similar universality has been shown for its
random-bond version~\cite{malakis09}. The $S=1$ model, studied in
this paper, is the 2d BC model at zero crystal field and therefore
it belongs to the same universality class with the 2d Ising model.
Of course, this universality should be expected to hold also for
all Archimedean lattices.

It is well known that the accuracy of MC data may be decisive for
a successful FSS estimation of critical properties. Over the
years, the numerical estimation of critical exponents has been a
non-trivial exercise, even for the simpler models, such as the
Ising model. An importance sampling approach, close to a
second-order phase transition, requires  appropriate use of
cluster algorithms that can efficiently overcome the well known
effects of critical slowing down. Wolff-type
algorithms~\cite{Swendsen87,Newman99,LandBind00} are easy to
implement and very efficient close to the critical point. The
Wolff algorithm will be implemented, in the present paper, to
simulate both the Ising and the $S=1$ model. However, for the
$S=1$ model the Wolff algorithm can not be used alone, because
Wolff steps act only on the non-zero spin values. A suggested
practice is now a hybrid algorithm along the lines followed by
Ref.~\cite{Blote95}. Since, we wanted to use a unified code for
both models the hybrid approach was tested and implemented for
both models. An elementary Monte carlo step of this scheme consist
of a number of Wolff steps (typically 5 Wolff-steps) followed by a
Metropolis sweep of the lattice. The combination with the
Metropolis lattice sweep is dictated by the fact that the Wolff
steps act only on the non-zero spin values.

Thus, we have simulated both the Ising model and the $S=1$ model
on the two ALs by implementing the same hybrid approach described
above. For the Ising model case, we carried out simulations only
at the exactly known critical temperatures~\cite{Codelo10},
whereas for the $S=1$ model we generated MC data to cover several
finite-size anomalies. In this case, the hybrid approach was
carried over to a certain temperature range depending on the
lattice size. Furthermore, for this case, we found it convenient
and of comparable accuracy to implement a parallel tempering (PT)
protocol, based on temperature sequences corresponding to an
exchange rate 0.5. This PT approach is very close to the practice
suggested recently in~\cite{bittner11}. The temperature sequences
were generated by short preliminary runs. Using such runs and a
simple histogram method~\cite{Swendsen87,Newman99}, the energy
probability density functions can be obtained and from these the
appropriate sequences of temperatures can be easily
determined~\cite{bittner11}.

\begin{figure}[htbp]
\includegraphics*[width=9 cm]{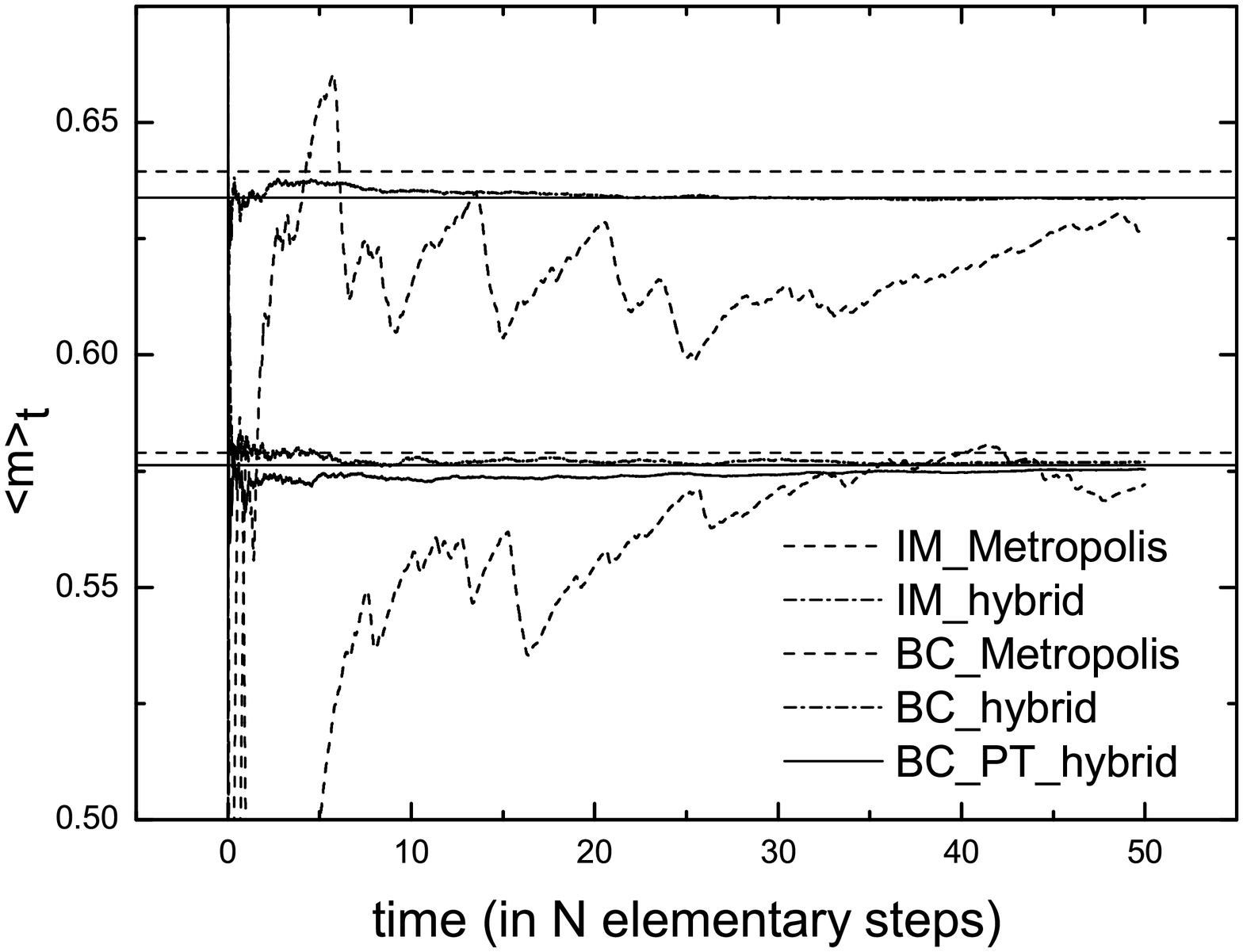}
\caption{\label{Fig:2} Behavior of moving averages for the order
parameter close to the critical temperatures for both the Ising
model (IM upper part of the graph), and the $S=1$ model (BC) on
the same (3,4,6,4) AL with $L=48$. The dashed and continuous
straight lines give averages over 20 independent runs for the
Metropolis algorithm (dashed lines), the hybrid and PT-hybrid
approach (continuous lines). The Metropolis algorithm has a very
slow equilibration. For the time unit the elementary steps as
defined in the text.}
\end{figure}

The superiority of the hybrid approach, over a simple Metropolis
scheme~\cite{metro53}, is illustrated in Fig.~\ref{Fig:2}. This
figure is constructed by using moving averages for the order
parameter (${\langle m \rangle}_{t}$) close to the corresponding
critical temperatures for both the Ising model (denoted in the
panel as IM) and the $S=1$ model (denoted in the panel as BC) on
the same (3,4,6,4) AL of linear size $L=48$ and $N=L^{2}$
vertices. As can be seen, from this illustration, the Metropolis
algorithm suffers from very strong fluctuations. It follows a very
slow approach to equilibrium and only with the help of heavy
sampling (the dashed lines give the average over 20 independent
Metropolis runs) its results are of reasonable accuracy. On the
other hand, the hybrid approach converges very fast to equilibrium
and produces accurate results even in only one single run. Here it
should be noted that, both the simple hybrid approach and its
combination with parallel tempering (in a convenient temperature
range) give in 20 independent runs the same results (with
comparable accuracy) and these are indicated by the continuous
straight lines in the panel of Fig.~\ref{Fig:2}.

Let us discuss now the FSS tools used in this paper for the
estimation of critical properties of the systems. In order to
estimate the critical temperature, we follow the practice of
simultaneous fitting approach of several pseudocritical
temperatures~\cite{malakis09}. From the MC data, several
pseudocritical temperatures are estimated, corresponding to
finite-size anomalies, and then a simultaneous fitting is
attempted to the expected power-law shift behavior
$T=T_{c}+bL^{-1/\nu}$. The traditionally used specific heat and
magnetic susceptibility peaks, as well as, the peaks corresponding
to the following logarithmic derivatives of the powers $n=1,2,4$
of the order parameter with respect to the inverse temperature
$K=1/T$~\cite{ferrenberg91},
\begin{equation}
\label{eq:4} \frac{\partial \ln \langle M^{n}\rangle}{\partial
K}=\frac{\langle M^{n}H\rangle}{\langle M^{n}\rangle}-\langle
H\rangle,
\end{equation}
and the peak corresponding to the absolute order-parameter
derivative
\begin{equation}
\label{eq:5} \frac{\partial \langle |M|\rangle}{\partial
K}=\langle |M|H\rangle-\langle |M|\rangle\langle H\rangle,
\end{equation}
will be implemented for a simultaneous fitting attempt of the
corresponding pseudocritical temperatures. Furthermore, the
behavior of the crossing temperatures of the 4th-order Binder
cumulants~\cite{binder81}, and their asymptotic trend, has been
observed and utilized for a safe estimation of the critical
temperatures.

The above described simultaneous fitting approach provides also an
estimate of the correlation length exponent $\nu$. An alternative
estimation of this exponent is obtained from the behavior of the
maxima of the logarithmic derivatives of the powers $n=1,2,4$ of
the order parameter with respect to the inverse temperature, since
these scale as $L^{1/\nu}$ with the system
size~\cite{ferrenberg91}. If the exponent $\nu$ has been
estimated, then the behavior of the values of the peaks
corresponding to the absolute order-parameter derivative, which
scale as $L^{(1-\beta)/\nu}$ with the system
size~\cite{ferrenberg91}, gives one route for the estimation of
the exponent ratio $\beta/\nu$. Further, knowing the exact
critical temperatures, or very good estimates of them, we can
utilize the behavior of the order parameter at the critical
temperatures for the traditional and effective estimation of the
exponent ratio $\beta/\nu$. Summarizing, our FSS approach
utilizes, besides the traditionally used specific heat and
magnetic susceptibility maxima, the above four additional
finite-size anomalies for the accurate estimation of the critical
temperature and critical exponents.

\section{The Ising model ($S=1/2$) on the (3,4,6,4) and ($3^4$,6) lattices}
\label{sec:3}

This Section presents the FSS analysis for the Ising model on the
two AL. The analysis is carried out only at the exactly known
critical temperatures. For the (3,4,6,4) lattice the exact
critical temperature is $T_{c}=2.1433...$~\cite{Codelo10} and for
each lattice size ($L=18,24,30,48,54,60,\ldots,138,144,150$) we
carried out 20 independent runs of the hybrid Metropolis-Wolff
algorithm at this temperature. The same number of independent runs
was carried out for the ($3^4$,6) lattice at the corresponding
exact critical temperature ($T_{c}=2.7858\ldots$~\cite{Codelo10}).
In this case, we have used a more dense sequence of lattice sizes
(a 6-step sequence): $L=18,24,30,36,\ldots,150$. We also give an
indication of the number of sweeps used in our final runs. For
each independent run we used for averaging $3\, 10^5$ sweeps for
the lattice with linear size $L=108$ and $5\, 10^5$ for the
lattice of size $L=150$. Equilibration periods were approximately
a third of the corresponding averaging time.
\begin{figure}[htbp]
\includegraphics*[width=8 cm]{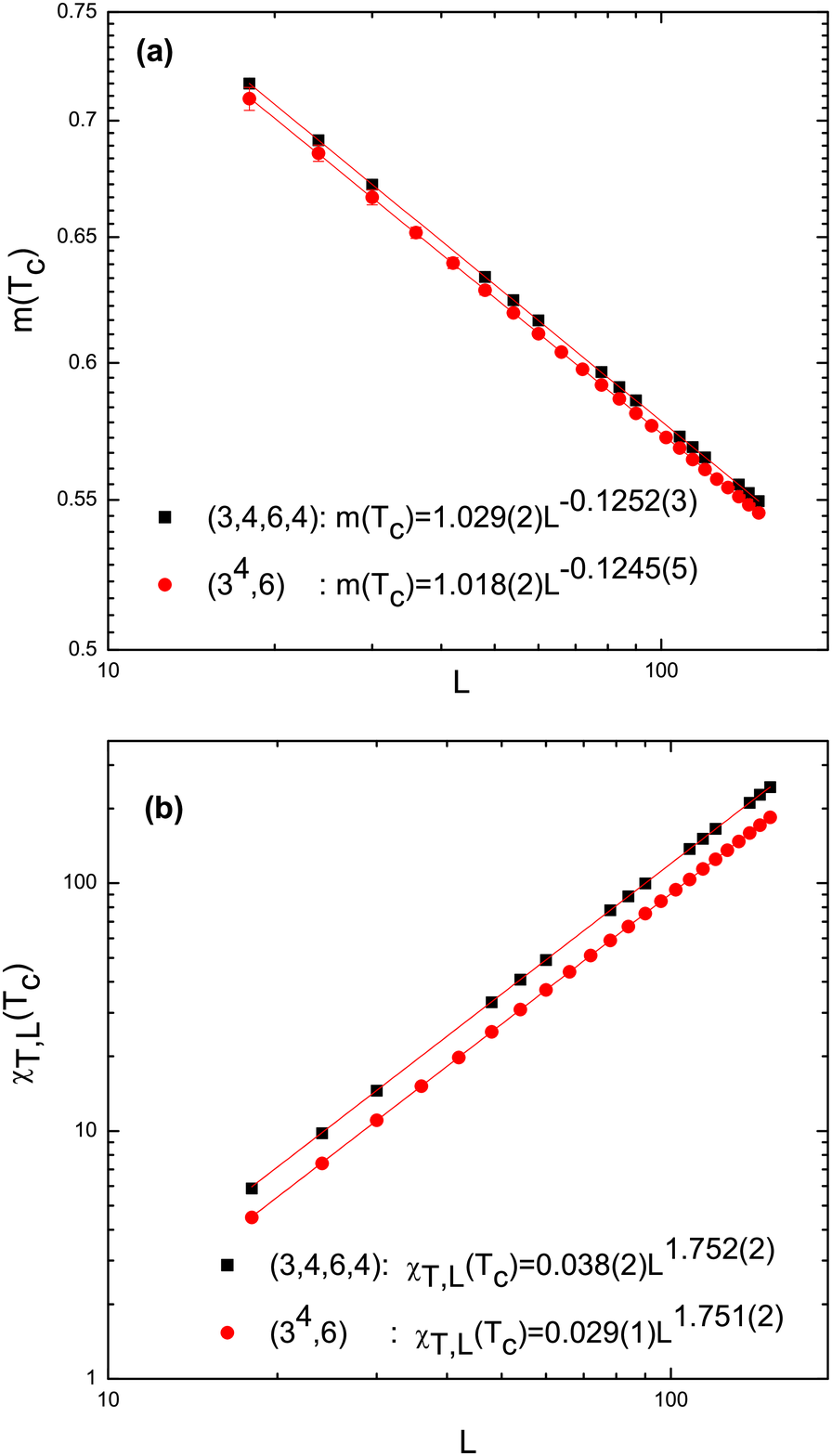}
\caption{\label{Fig:3} (Color online) (a) Finite-size behavior of
the order-parameter at the exact critical temperatures,
illustrated in a logarithmic scale, for both ALs. In the panel we
show a simple power law estimation for the exponent ratio
$\beta/\nu$. (b) The same for the FSS behavior of the
susceptibility. In the panel we show a simple power law estimation
for the exponent ratio $\gamma/\nu$. }
\end{figure}

We start the presentation of our FSS attempts by illustrating the
behavior of the order parameter at the exact critical temperatures
on the two AL. This behavior is illustrated, in a logarithmic
scale, in Fig.~\ref{Fig:3}(a). In the panel of this figure we show
a simple power-law estimation for the exponent ratio $\beta/\nu$.
This simple estimation gives an accuracy to the third significant
figure of the exact critical exponent ratio $\beta/\nu=0.125$. We
point out that, the fitting parameters are not sensitive to
fitting lattice-range used ($L=18-150$) and are almost identical
if we do not include, in the fitting attempts, the statistical
errors shown in the panel. The errors bars shown, were calculated
by the jackknife method for each run. The estimated errors for 20
independent runs are shown in panel (a) and were used in the
fitting attempt. For all other (diverging) thermodynamic
parameters, such as the susceptibility, the corresponding
jackknife errors are again very small, smaller than the symbol
sizes, and are therefore omitted in the sequel. We continue by
presenting now the estimation of the exponent ratio that
characterizes the divergence of the susceptibility at the critical
temperature. This behavior is illustrated, again in a logarithmic
scale, in Fig.~\ref{Fig:3}(b). In the panel of this figure we show
a simple power-law estimation for the exponent ratio $\gamma/\nu$.
For both lattices the simple power law gives again an accuracy to
the third significant figure of the exact critical exponent ratio
$\gamma/\nu=1.75$.

\begin{figure}[htbp]
\includegraphics*[width=8 cm]{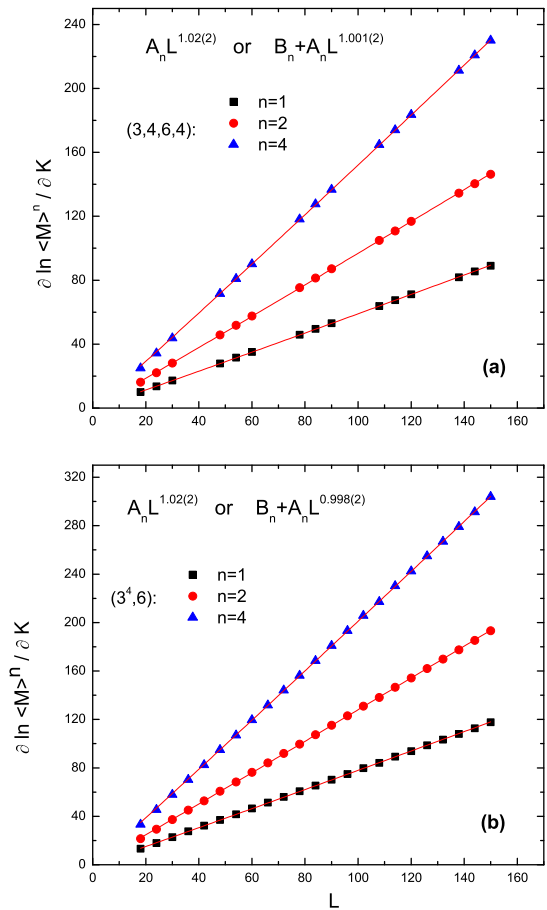}
\caption{\label{Fig:4} (Color online) (a) FSS behavior of the
logarithmic derivatives of the powers $n=1,2,4$ of the order
parameter with respect to the inverse temperature for the
(3,4,6,4) AL. Simultaneous fitting attempts to a simple power law
and to a power law with a constant correction term are shown in
the panel. (b) The same as (a) for the ($3^4$,6) lattice.}
\end{figure}

The critical exponent of the correlation length can be estimated
from the behavior of the logarithmic derivatives of the powers
$n=1,2,4$ of the order parameter with respect to the inverse
temperature. As pointed out earlier, these scale as $L^{1/\nu}$
with the system size~\cite{ferrenberg91} and their behavior
provides an alternative route for the estimation of the
correlation length critical exponent. Their behavior is
illustrated in Figs.~\ref{Fig:4}(a) and (b) respectively for the
two ALs. Our practice here, is to use a simultaneous fitting
attempt to a simple power law for the three cases $n=1,2,4$ in
each lattice. In the panel of these figures we show that a simple
power-law estimation provides the estimates $1/\nu=1.02(2)$ for
both lattices. However, we point out that the fitting attempts are
significantly improved here if we include suitable correction
terms. One possibility is to include a constant term which does
not, of course, effects the divergence of the susceptibility at
the critical temperatures. As shown in the panels the estimates
are now $1/\nu=1.001(2)$ for the (3,4,6,4) and $1/\nu=0.998(2)$
for the ($3^4$,6) AL, giving strong and clear evidence of $\nu=1$.

\begin{figure}[htbp]
\includegraphics*[width=9 cm]{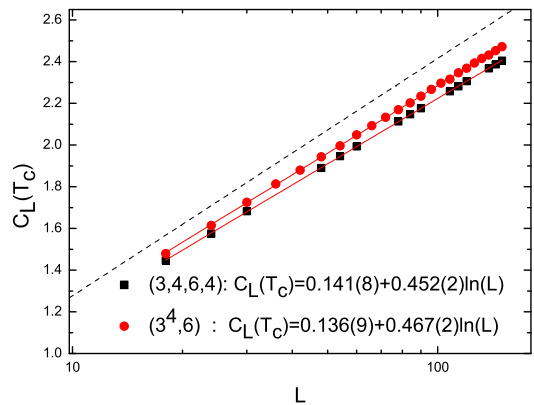}
\caption{\label{Fig:5} (Color online) FSS behavior of the specific
heat at the exact critical temperatures for the two ALs,
illustrated in semi-logarithmic scale. This behavior is described
in detail in the text and is compared here with the dashed line
which describes the FSS behavior of the specific heat, at the
exact critical temperature, of the Ising model on the square
lattice with periodic boundary conditions.}
\end{figure}

We close this section by discussing the FSS of the specific heat
at the exact critical temperatures for the two ALs. From the work
of Ferdinand and Fisher~\cite{Ferdin69} we know the characteristic
specific heat expansion for the square lattice Ising model.
Details of the size expansions on this lattice, but also on some
other 2d lattices (plane triangular and honeycomb lattices) have
been published in a number of
papers~\cite{malakis04,Ferdin69,Izmail02,Izmail03,salas01}. This
is an interesting topic and one should expect that similar
expansions are to be obeyed for all ALs. In Fig.~\ref{Fig:5} we
illustrate the expected logarithmic divergence of the specific
heat at the exact critical temperatures for the two ALs. Our
fitting attempts have been restricted to the leading behavior
$C_{L}(T_{c})=B_{c}+A_{0}\ln(L)$, which avoids expected higher
order ($L^{-1}$, $L^{-2}$, $\ldots$) correction
terms~\cite{salas01}. This practice sidesteps problems of
competition between small, but unavoidable, statistical errors and
small correction terms. In the panel of Fig.~\ref{Fig:5} we show
the estimated critical amplitudes $A_{0}$ for the two ALs studied
here and also the constant $B_{c}$ contributions. The estimation
has been done by using the full size range $L=18-150$. However, as
mentioned, because of statistical errors and small higher-order
corrections these estimations show some sensitivity to the
size-range used. Observing the asymptotic trend, our best
estimates for the critical amplitudes are $A_{0}=0.450(8)$ for the
(3,4,6,4) lattice and $A_{0}=0.464(8)$ for the ($3^4$,6) lattice.
The constant contributions are more sensitive and our moderate
estimates are of the order of $B_{c}=0.15(3)$ for both ALs. For
comparison the leading behavior for the square lattice Ising
model, $C_{L}(T_{c})=0.138149\ldots+0.494538\ldots \ln(L)$, is
illustrated in the same figure by the dashed line. We point out
that the small distance between the estimates for the critical
amplitudes of the two ALs should not be taken as a sign indicating
a possible equality. For instance a similar situation can be found
between the square and plane triangular lattices with amplitudes
$A_{0}=0.494538\ldots$ and $A_{0}=0.499069\ldots$
respectively~\cite{Izmail03}. In conclusion, as expected, all
critical properties of the 2d Ising model, critical exponents and
critical expansions, are well obeyed on the two Archimedean
lattices studied here.

\begin{figure}[htbp]
\includegraphics*[width=8 cm]{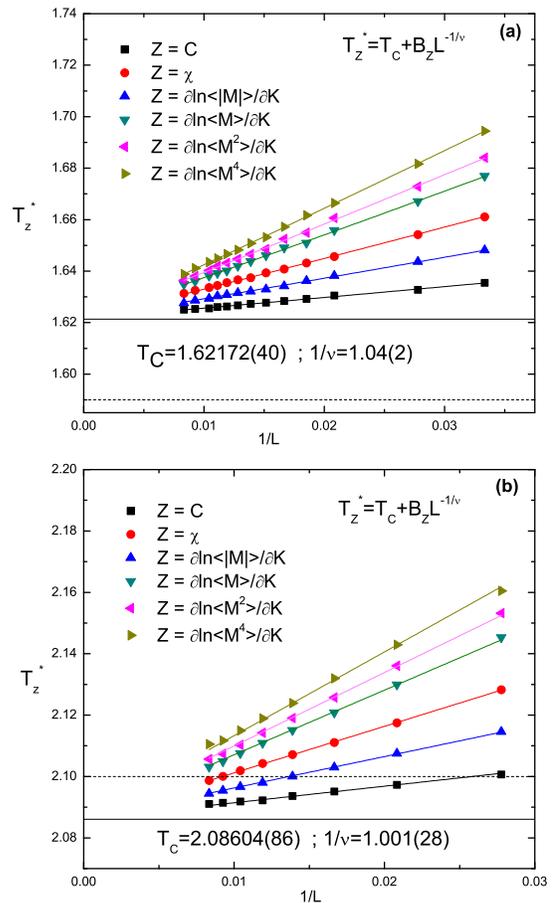}
\caption{\label{Fig:6} (Color online) FSS behavior of the six
pseudocritical temperatures defined in the text for the two ALs
(panels (a) and (b) for the (3,4,6,4) and ($3^4$,6) respectively)
for the $S=1$ model. Corresponding estimates for the critical
temperatures and $1/\nu$, of the illustrated fitting attempts, are
given in the panels and further discussed in the text. Our final
estimates of the critical temperatures (see the text) are shown by
the continuous straight lines in the corresponding panels, whereas
dashed straight lines indicate the estimates for the critical
temperatures of Ref.~\cite{Lima10}.}
\end{figure}

\section{The $S=1$ model on the (3,4,6,4) and ($3^4$,6) lattices }
\label{sec:4}

This Section presents the critical properties of the $S=1$ model
on the (3,4,6,4) and ($3^4$,6) ALs. The MC data were generated by
the combination of the hybrid approach with the PT protocol,
described in Sec.~\ref{sec:2}, and we have averaged over five
independent runs, in the appropriate temperature ranges, and use
linear sizes $L=30,36,48,54,60,66,72,78,84,90,96,108,120$ for the
(3,4,6,4) lattice and $L=36,48,60,72,84,96,108,120$ for the
($3^4$,6) lattice. As discussed in Sec.~\ref{sec:2}, the
second-order transition of this model between the ferromagnetic
and paramagnetic phases is expected to be in the universality
class of the simple 2d Ising model. We will verify this
expectation and contrast our findings with those in the report of
Ref.~\cite{Lima10}.

Figure~\ref{Fig:6} presents the shift behavior of several
pseudocritical temperatures for the $S=1$ model on the two ALs
(panels (a) and (b)). These temperatures correspond to the peaks
of the following six quantities: specific heat, magnetic
susceptibility, inverse temperature derivative of the absolute
order parameter, and inverse temperature logarithmic derivatives
of the first, second, and fourth powers of the order parameter.
The data are fitted, in the corresponding size ranges ($L=30-120$
and $L=36-120$) to the expected power-law behavior
$T=T_{c}+bL^{-1/\nu}$ and the resulting estimates of the critical
temperatures and $1/\nu$ are given in the panels. To some degree
these estimates are sensitive to the size range used and to
statistical errors. However, by varying the size ranges and also
observing the asymptotic trend of the crossing temperatures of the
4th order Binder cumulants we have with confidence estimate that
the critical temperatures are $T_{c}=1.62115(55)$ for the
(3,4,6,4) lattice and $T_{c}=2.08605(15)$ for the ($3^4$,6)
lattice. These values are indicated by the continuous straight
lines in the corresponding panels. With dashed straight lines we
have also indicated the respective estimates for the critical
temperatures $T_{c}=1.590(3)$ and $T_{c}=2.100(3)$ of
Ref.~\cite{Lima10}.

\begin{figure}[htbp]
\includegraphics*[width=8 cm]{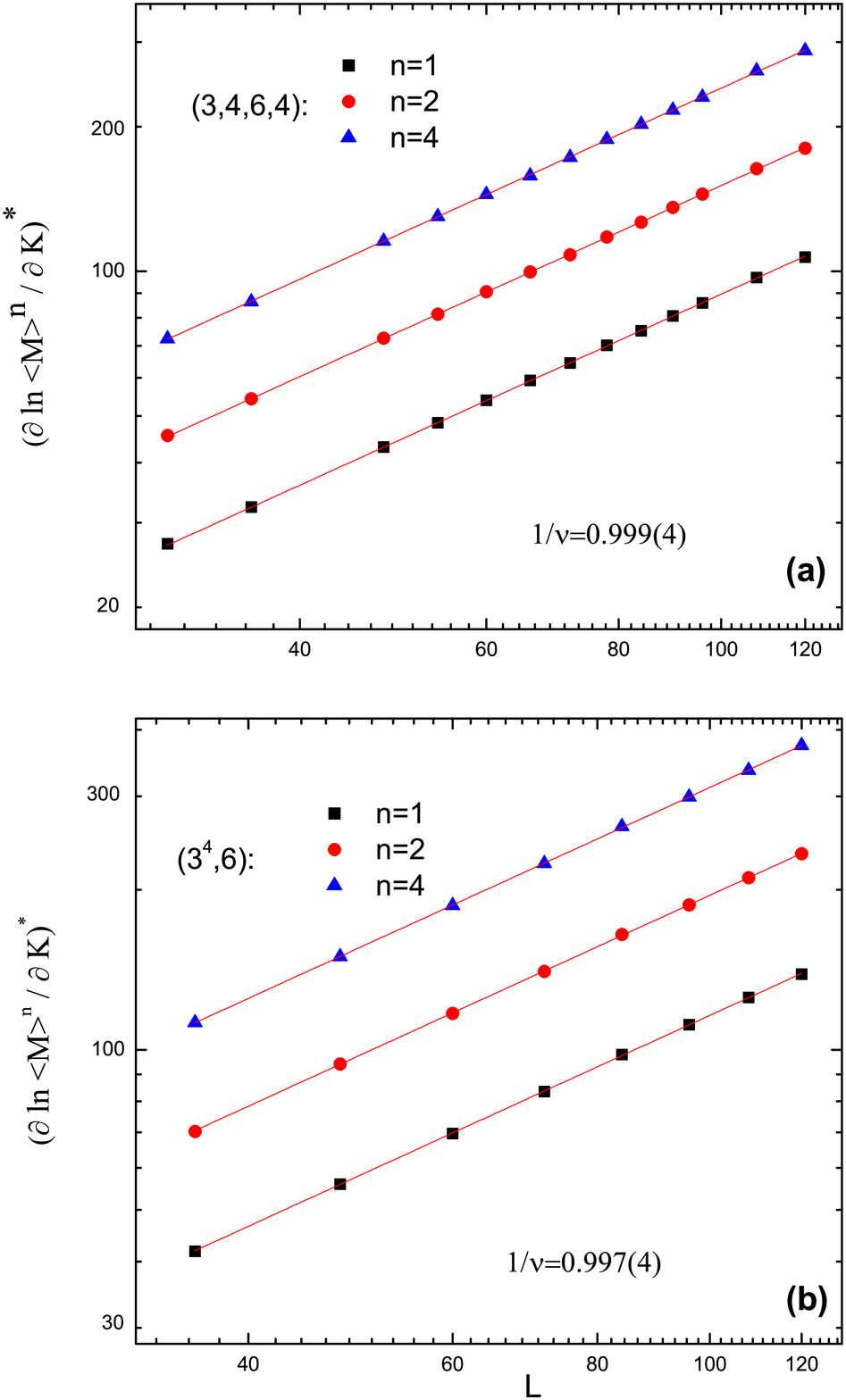}
\caption{\label{Fig:7} (Color online) FSS behavior, illustrated in
a logarithmic scale, of the peaks of the logarithmic derivatives
of the powers $n=1,2,4$ of the order parameter with respect to the
inverse temperature for the two ALs for the $S=1$ model.
Corresponding estimates for the exponent $1/\nu$ are given in the
panels by applying simultaneous fitting attempts to a simple power
law.}
\end{figure}

\begin{figure}[htbp]
\includegraphics*[width=8 cm]{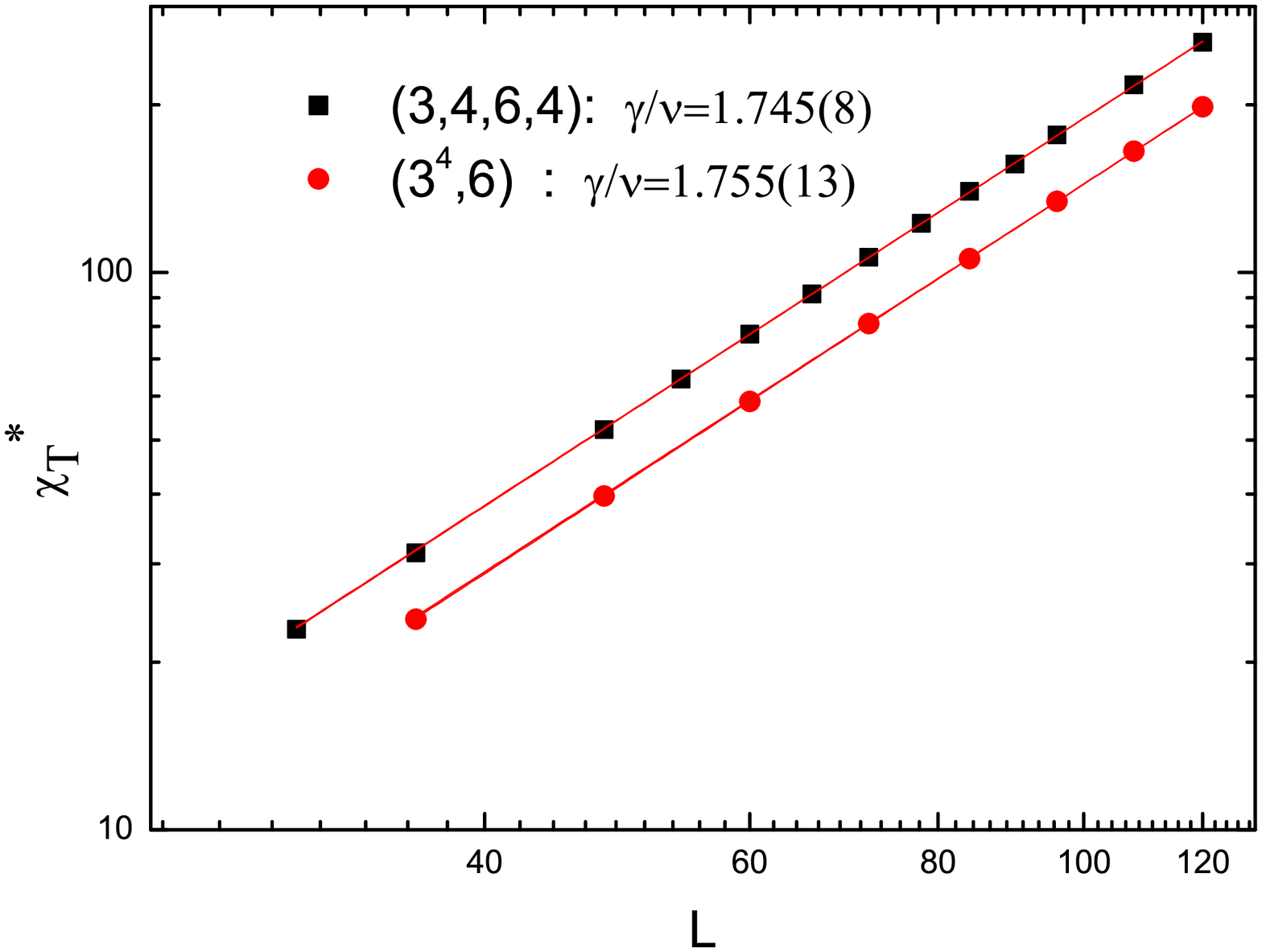}
\caption{\label{Fig:8} (Color online) FSS behavior of the magnetic
susceptibility maxima illustrated in a logarithmic scale for both
ALs for the $S=1$ model. In the panel we show simple power-law
estimations for the exponent ratio $\gamma/\nu$.}
\end{figure}

\begin{figure}[htbp]
\includegraphics*[width=8 cm]{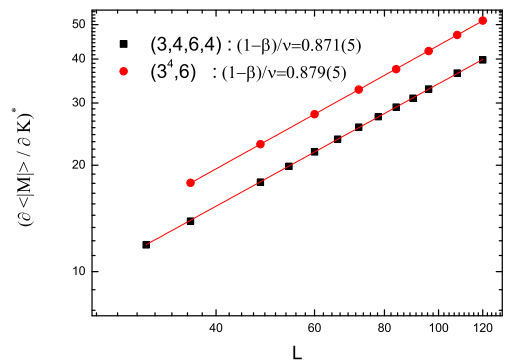}
\caption{\label{Fig:9} (Color online) Estimations for the magnetic
exponent ratio $\beta/\nu$ for both ALs for the $S=1$ model,
obtained from the analysis of the corresponding finite-size
anomalies (absolute order-parameter derivative). In the panel we
show simple power-law estimations for this exponent.}
\end{figure}

Their estimates for $T_{c}$ are in serious errors and so are the
values $1/\nu=0.83(5)$ and $1/\nu=0.94(5)$ for the correlation
length exponent reported in their paper~\cite{Lima10}. On the
other hand, our estimates for this exponent, as shown in the
panels, are clear indications of the universality mentioned above.
A further verification for this is the behavior of the logarithmic
derivatives of the powers $n=1,2,4$ of the order parameter with
respect to the inverse temperature. Their behavior is illustrated
in Fig.~\ref{Fig:7} (a) and (b) respectively for the two ALs. The
estimates from the simultaneous fitting attempt are shown in the
corresponding panels. Figure~\ref{Fig:8} and Fig.~\ref{Fig:9},
present our estimations for the magnetic exponent ratios
$\gamma/\nu$ and $\beta/\nu$, obtained from the analysis of the
corresponding finite-size anomalies (peaks). Again, they provide
very strong verification of the expected universality. In
conclusion, our results for the 2d BC model at $\Delta=0$ are in
full agreement with the universality arguments that place the BC
model in the Ising universality class.

\begin{figure}[htbp]
\includegraphics*[width=8 cm]{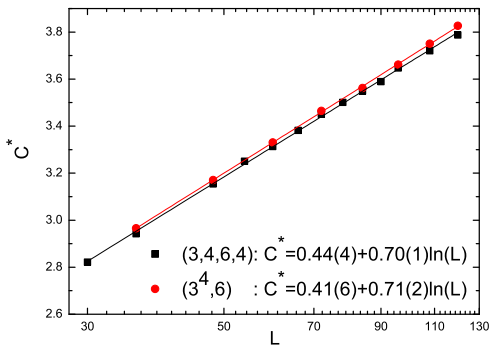}
\caption{\label{Fig:10} (Color online) FSS behavior of the
specific heat at the specific heat's pseudocritical temperatures
(specific heat peaks) for the two ALs, for the $S=1$ model,
illustrated in semi-logarithmic scale. This behavior is further
discussed in the text.}
\end{figure}

\begin{figure}[htbp]
\includegraphics*[width=8 cm]{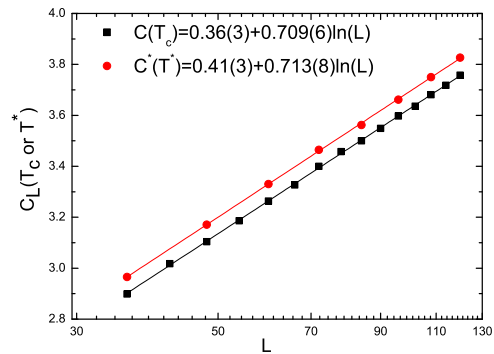}
\caption{\label{Fig:11} (Color online) FSS behavior of the
specific heat at the specific heat's pseudocritical temperature
and at the estimated exact critical temperature for the ($3^4$,6)
AL, for the $S=1$ model, illustrated in semi-logarithmic scale.
The critical amplitude is further discussed in the text.}
\end{figure}

We now turn to interesting topic of estimating the critical
amplitudes of the logarithmic specific heat expansions for the
$S=1$ model on the (3,4,6,4) and ($3^4$,6) AL. In
Fig.~\ref{Fig:10} we plot the expected logarithmic divergences of
the specific heat at the specific heat's pseudocritical
temperatures for the two lattices. Again, and for similar reasons,
our fitting attempts are restricted to the leading behavior
$C^{\star}=B^*+A_{0}ln(L)$. The estimates for the critical
amplitudes, given in the panels, are very close to each other for
the two ALs, but as mentioned earlier there are not any reasons to
expect their equality. Since the estimate for the critical
temperature ($T_{c}=2.08605(15)$) for the ($3^4$,6) AL appears to
be accurate to at least five significant figures, we have
undertake for this lattice $20$ independent runs using the hybrid
approach only at this temperature. Figure~\ref{Fig:11} illustrates
and contrast the expected logarithmic divergences of the specific
heat at the specific heat's pseudocritical temperature and at the
critical temperature $T_{c}=2.08605(15)$ for the $S=1$ model on
the ($3^4$,6) AL. It is notable here that, the estimations in the
panels for the critical amplitude are obtained by applying two
independent fits, whereas a simultaneous fitting attempt gives
$A_{0}=0.7103(46)$. This appears to be an accurate estimation and
the values in the panels (which should be equal) are both within
its error limits. This critical amplitude for the $S=1$ model can
be compared with the corresponding value $A_{0}=0.464(8)$ for the
Ising model on the same ($3^4$,6) Archimedean lattice. A similar
project was carried out for the (3,4,6,4) AL and our best estimate
for the critical amplitude is $A_{0}=0.700(9)$, which now should
be compared to the corresponding value $A_{0}=0.450(8)$ for the
Ising model on the same AL.
Let us close this Section by pointing out that from the MC data,
at the estimated critical temperatures for the $S=1$ model on the
two ALs, we also carried out the estimation of all critical
exponents, by using the FSS tools mentioned in Sec.~\ref{sec:2}.
All obtained estimates were in excellent agreement verifying not
only the expected universality but also the accuracy of the
estimated critical temperatures.

\section{Conclusions}
\label{sec:5}

The Ising $S=1/2$ and the $S=1$ models have been studied on two
Archimedean lattices by an efficient Monte Carlo scheme, using a
hybrid Wolff-Metropolis approach. The Ising model was analyzed by
finite-size scaling at the exact critical temperatures. We
verified, with high accuracy, all critical exponents of the well
known 2d Ising model exact values. For the $S=1$ model, on the
same lattices, we presented very clear evidence that this model
obeys, also very well, the 2d Ising model critical exponents. Our
results are in full agreement with the general universality
arguments that place these models on all 2d lattices in the 2d
Ising universality class. In conclusion, we have disclosed any
questions raised by the recent attempt~\cite{Lima10} to estimate
critical exponents on these lattices, for the $S=1$ model, and
define effective dimensionality. Their results, most likely,
suffer from strong critical slowing down effects, due to the
simple heat bath algorithm implemented by these authors. In
addition, we have provided reliable results for the characteristic
specific heat expansions on the two Archimedean lattices studied
here for both models.

\begin{acknowledgments}
The authors acknowledge useful discussions with Professor A.N.
Berker and his comments on the manuscript.
\end{acknowledgments}

{}

\end{document}